\documentstyle[aps,preprint,prb]{revtex}
\begin{document} 
\title {\bf Compensation of Magnetic Impurity Effect in Superconductors
\\ by Radiation Damage }
\author{ Mi-Ae Park$^{a,1}$, M. H. Lee$^{a}$, and Yong-Jihn Kim$^{b, *}$}
\address{ $^{a}$ Department of Physics, Seoul National University,
Seoul 151-742, Korea\\
$^{b}$Department of Physics, Korea Advanced Institute of Science and Technology,\\
Taejon 305-701, Korea}
\maketitle
\vspace{5pc}
\def\esn{\epsilon} 
\def\upa{\uparrow}
\def\dna{\downarrow}
\def\vq{\vec q} 
\def\vk{\vec k} 
\def\vr{\vec r}
\def\br{{\bf r}}
\def\bR{\bf R}
\def\bs{\bf s}
\def\bS{\bf S}
\def\bl{{\bf l}}
\def\hb{\hfill\break}
\vskip 2pc
Compensation of the reduction of $T_{c}$ caused by magnetic impurities
has been observed as a consequence of radiation damage.
Using the recent theory by Kim and Overhauser (KO) we consider the effect of
radiation damage on the $T_{c}$ of superconductors having magnetic impurities.
We find a good fitting to the experimental data. 
It is also pointed out that Gor'kov's
formalism with the pairing constraint derived from  
the Anomalous Green's function leads to KO theory.
 

\hspace{2pc}       Gor'kov's formalism, pairing constraint

\vspace{6pc}
$^{1}$ Present address: Department of Physics Education, Seoul National University,

\hspace{2pc} Seoul 151-742, Korea

$^{*}$ Corresponding author. URL: http://taesan.kaist.ac.kr/$\tilde{\ }$yjkim                
\vfill\eject  
 
\section*{1. Introduction}
\vspace{1pc}
Recently Kim and Overhauser (KO)$^{1}$ obtained different results 
for 
the magnetic impurity effect on superconductors compared with those of  
Abrikosov and Gor'kov's theory.$^{2}$ 
First, the initial slope of $T_{c}$ decrease by magnetic impurities is found
to depend on the superconductor and therefore is not the universal
constant proposed by Abrikosov and Gor'kov.
Second, the reduction of $T_{c}$ by magnetic impurities is significantly
lessened whenever the mean free path $\ell$ becomes smaller than the 
BCS coherence length $\xi_{o}$. This compensation phenomenon has been
observed by adding non-magnetic impurities$^{3-5}$ and radiation 
damage,$^{6-8}$
whereas prior theories predict that
magnetic impurity effect is not influenced by the non-magnetic scattering.

In this paper we compare the theoretical $T_{c}$ values calculated by KO theory
with the data of Hofmann, Bauriedl, and Ziemann.$^{8}$
Fairly good agreement was found.
Hofmann et al. irradiated pure In and In + 400 ppm Mn foils with Ar ions.
A $\Delta T_{c}= 2.2K$ in $T_{c}$ for a pure 70 nm In film compared to an
identical film ion implanted with 400 ppm Mn was changed to
$\Delta T_{c}=0.3K$ after both films were exposed to a 275 kev $Ar^{+}$-ion fluence
of $2.2\times 10^{16}cm^{-2}$. Both films were maintained 
below 15K during the $Ar^{+}$ irradiation.
The reason of this compensation phenomenon is that only magnetic
solutes within $\xi_{eff} \sim (\ell\xi_{o})^{1\over 2}$ of a Cooper
pair's center of mass can diminish the pairing interaction. 

We also point out the problem inherent in the self-consistency equation of the
Gor'kov's formalism.$^{9,10}$ 
In the presence of the magnetic impurities the self-consistency equation
fails to choose a correct pairing, which is consistent with the
physical constraint of the system.
The self-consistency equation allows  some extra pairing terms forbidden by the
physical constraint. 
The remedy is the following: we first find a 
 correct form of the Anomalous Green's function satisfying the physical 
constraint and then derive a self-consistency equation from it.
In that case the revised self-consistency equation gives nothing but
Kim and Overhauser's result.$^{1}$ 

\section*{2. BCS-type Theory by Kim and Overhauser}
\vspace{1pc}
We will briefly review KO's approach.$^{1}$
The magnetic interaction between a conduction electron at $\br$ and a magnetic 
impurity (having spin $\bS$), located at ${\bR}_{i}$, is given by
\begin{equation}
H_{m}({\br})=J{\bs}\cdot{\bS}_{i}v_{o}\delta({\br}-{\bR}_{i}),
\end{equation} 
where ${\bs}={1\over 2}\sigma$ and $v_{o}$ is the atomic volume.
In the presence of the magnetic impurities BCS pairing must employ degenerate
partners which have the $J{\bs}\cdot{\bS}_{i}$ scattering built in
because the strength of exchange scattering $J$ is much larger than
the binding energy. This scattered state representation was first introduced
by Anderson in his theory of dirty superconductors.$^{11}$
 The scattered basis state which carries the 
label, $\vk\alpha$, is 
\begin{equation}
\psi_{\vk\alpha} = N_{\vk}\Omega^{-{1 \over 2}} [ e^{i\vk\cdot \vr}\alpha
+ \sum_{\vq}e^{i(\vk + \vq)\cdot \vr}(W_{\vk\vq}\beta + W_{\vk\vq}^{'}
\alpha)], 
\end{equation}
where,  
\begin{equation}
W_{\vk\vq} = {{1 \over 2}J\overline{S}v_{o}\Omega^{-1} \over 
\epsilon_{\vk} - \epsilon_{\vk+\vq}}\sum_{j}sin \chi_{j} e^{i\phi_{j}
-i\vq\cdot {\bR}_{j} }
\end{equation}
and,
\begin{equation}
W_{\vk\vq}' = {{1 \over 2}J\overline{S}v_{o}\Omega^{-1} \over
\epsilon_{\vk} - \epsilon_{\vk + \vq}}\sum_{j} cos \chi_{j} e^{-i\vq\cdot
{\bR}_{j}}. 
\end{equation}
$\chi_{j}$ and $\phi_{j}$ are the polar and azimuthal angles of the spin
${\bS}_{j}$ at ${\bR}_{j}$, and the $\epsilon$'s are the electron energies
of the host. The perturbed basis state for the degenerate partner of (2)
is:
\begin{equation}
\psi_{-\vk\beta} = N_{\vk}\Omega^{-{1 \over 2}}[e^{-i\vk\cdot\vr}\beta
+ \sum_{\vq}e^{-i(\vk + \vq)\cdot \vr} (W_{\vk\vq}^{*}\alpha - W_{\vk\vq}
^{'*}\beta)]. 
\end{equation}

At each point $\vr$, the two spins of the degenerate partner become
canted by the mixing of the plane wave and spherical-wavelet
component. Consequently, the BCS condensate is forced to have a triplet
component because of the canting caused by the exchange scattering.
The new matrix element between the canted basis pairs is (to order $J^{2}$)
\begin{equation}
V_{\vk'\vk} = - V < cos\theta_{\vk'}(\vr)> < cos\theta_{\vk}(\vr)>,
\end{equation}
where $\theta$ is the canting angle.
The angular brackets indicate both a spatial and impurity average.
It is given
\begin{equation}
<cos\theta_{\vk}(\vr)>\ \cong \ 1 -2|W_{\vk}|^{2}, 
\end{equation}
where $|W_{\vk}|^{2}$ is the relative probability contained in the
virtual spherical waves surrounding the magnetic solutes (compared
to the plane-wave part).
From Eqs. (2)-(4) we obtain
\begin{equation}
|W_{\vk}|^{2} = {J^{2}m^{2}{\bar S}^{2}c_{m}R\over 8\pi n\hbar^{4}}.
\end{equation}
Because the pair-correlation amplitude 
falls exponentially as $exp(-r/\pi\xi_{o})$$^{12}$ at $T=0$ and
as $exp(-r/3.5\xi_{o})$$^{13}$ near $T_{c}$, 
we set
\begin{equation}
R = {3.5\over 2}\xi_{o}. 
\end{equation}
Then one finds
\begin{equation}
<cos\theta> = 1 - {3.5\xi_{o} \over 2 \ell_{s}}, 
\end{equation}
where $\ell_{s} = v_{F}\tau_{s}$ is the mean free path for exchange 
scattering only.

 The BCS $T_{c}$
equation still applies after a modification of the effective coupling
constant according to Eq. (6):
\begin{equation}
\lambda_{eff} = \lambda <cos\theta>^{2}, 
\end{equation}
where the BCS $\lambda $ is $N_{o}V.$ Accordingly, the BCS $T_{c}$ equation
is now,
\begin{equation}
k_{B}T_{c} = 1.13\hbar\omega_{D}e^{-{1\over \lambda_{eff}}}. 
\end{equation}
The initial slope is given
\begin{equation}
k_{B}(\Delta T_{c}) \cong -{0.63\hbar \over \lambda\tau_{s}}. 
\end{equation}
The factor $1/\lambda$ shows that the initial slope depends on the
superconductor and is not  a universal constant.
For an extended range of solute concentration,
KO find 
\begin{equation}
<cos\theta> = {1\over 2} + {1\over 2}[1 + 5({u\over 2})^{2}]^{-1}
e^{-2u}, 
\end{equation}
where
\begin{equation}
u \equiv 3.5\xi_{eff}/2\ell_{s}. 
\end{equation}

When the conduction electrons have a mean free path $\ell$ which is smaller
than the coherence length $\xi_{o}$ (for a pure superconductor), the
effective coherence length is 
\begin{equation}
\xi_{eff} \approx \sqrt{\ell\xi_{o}}. 
\end{equation}
For a superconductor which has ordinary impurities as well
as magnetic impurities, the total mean-free path $\ell$ is given by 
\begin{equation}
{1\over \ell} = {1\over \ell_{s}} + {1\over \ell_{o}}, 
\end{equation}
where $\ell_{o}$ is the potential scattering mean free path.
It is clear from Eq. (16) that the potential scattering 
profoundly affects the paramagnetic impurity effect. 
In other words, the size of the Cooper pair is reduced by the potential
scattering and the reduced Cooper pair sees a smaller number of
magnetic impurities. Accordingly the magnetic impurity effect is
partially suppressed.  This is the origin of the compensation phenomena 
observed in experiments.$^{3-8}$

\section*{3. Comparison with Experiments}
\vspace{1pc}
Now we compare the KO theory with experiment.
In Fig. 1 the $T_{c}$ of In (open symbols) and InMn (closed symbols) 
were plotted as a function of $Ar^{+}$ fluence.
The data are due to Hofmann, Bauriedl, and Ziemann.$^{8}$
Well annealed In-Films with the thickness of 70nm were prepared.
The mean value of the residual resistivities, $\rho_{i}$, was 0.62$\mu\Omega$
with a variation of 20\%.
During the irradiation, In-films were maintained below 15K.
As you see, irradiation induces the increase of the transition temperature 
of In-film, which may be due to the increase of electron-phonon 
interaction.
Open symbols were fitted by the BCS $T_{c}$ equation with 
\begin{equation}
\lambda=0.284\times tanh(0.55\Phi + 1.76).
\end{equation}
$\Phi$ denotes the $Ar^{+}$ fluence/$10^{16}$.
The Debye frequency $\omega_{D}$ of In was set to be 129K.
Closed symbols show the transition temperature of In-Mn alloys which 
were irradiated with $Ar^{+}$ after the Mn-implantation.
400PPM of Mn was implanted and led to $\Delta T_{c}\approx 2.2K$. 
Notice that $Ar^{+}$ irradiation not only increases the $T_{c}$ as in
the case of In-film but also suppress the $T_{c}$ decrease caused by 
Mn implantation.
This compensation of magnetic impurity effect by radiation damage 
contradicts Abrikosov-Gor'kov's theory. 
Merriam et al.$^{3}$ also found that adding dilute concentrations of
ordinary impurities such as Pb or Sn to bulk In samples significantly 
reduces the magnetic impurity effect. 

As we saw in Sec. 2, the ratio of the effective coherence length
to the spin disorder scattering length, $\xi_{eff}/\ell_{s}\approx
\sqrt{\ell\xi_{o}}/\ell_{s}$, determines $T_{c}$.
Using the Drude formula, $\rho=m/ne^{2}\tau$, 
we can calculate the electron mean free path $\ell=v_{F}\tau$.
For In $n=1.15\times 10^{23}cm^{-3}$ and 
$v_{F}=1.74\times 10^{8}cm/sec$.
The residual resistivity increase due to the 
Mn-implantation is estimated to be
$\Delta\rho_{Mn} \approx 1\mu\Omega cm$. 
On the other hand, the residual resistivity increase, $\Delta \rho_{Ar}$, 
due to Ar irradiation, 
measured by by Hofmann, Ziemann, and Buckel,$^{14}$ was fitted by
\begin{equation}
\Delta\rho_{Ar} =  6.5ln(\Phi + 1). 
\end{equation}
Consequently, the total resistivity, $\rho_{tot}$,  is 
\begin{eqnarray}
\rho_{tot} 
&=& \rho_{i} + \Delta\rho_{Ar} +  \Delta\rho_{Mn} \nonumber \\
&=& 0.62 + 6.5ln(\Phi + 1) + 1.0. 
\end{eqnarray}
From the total resistivity we can calculate $\ell$ and the effective coherence
length. With the effective coherence length we readily find $T_{c}$ by the
BCS $T_{c}$ equation. The theoretical curve shown (lower solid curve)
involves just one adjustable parameter, $\tau_{s}$, in order that
$T_{co} = 1.15K$, the observed value without irradiation. 
We used $\ell_{s} = 35330\AA$.
Because the In films are actually quasi-two dimensional,  
there may be some corrections
due to the finite thickness, which seem to be negligible in dirty
limit. 
Nevertheless the agreement is fairly good considering some uncertainties in 
the film thickness effect and in the calculation of the total resistivity. 

\section*{4. Gor'kov's Formalism with Pairing Constraint}
\vspace{1pc}
This compensation phenomenon
contradicts prior theories for magnetic solutes.$^{2}$
The failure of Abrikosov and Gor'kov's theory originates from the inclusion
of the {\sl extra pairing} terms which violate the physical constraint
of the Anomalous Green's function $F({\br}, {\br'})$.$^{9,10,15}$
Now we show how we can obtain the result of KO theory from the Gor'kov's
formalism.
For simplicity let's consider only the (spin-nonflip) z-component of 
the magnetic interaction.
Gor'kov's self-consistency equation is given
\begin{equation}
\Delta({\br}) = VT\sum_{\omega}\int \Delta({\bl})G^{\upa}_{\omega}({\br,\bl})
G^{\dna}_{-\omega}({\br,\bl})d\bl,
\end{equation}
where
\begin{equation}
G^{\upa}_{\omega}({\br,\bl}) = \sum_{\vk}{\psi_{\vk\upa}({\br})\psi_{\vk\upa}^{*}({\bl})
\over i\omega - \esn_{\vk}},  
\end{equation}
and
\begin{equation}
G^{\dna}_{-\omega}({\br',\bl}) = \sum_{\vk'}{\psi_{\vk'\dna}({\br'})
\psi_{\vk'\dna}^{*}({\bl})
\over -i\omega - \esn_{\vk'}}.  
\end{equation}
Note that $\psi_{\vk\upa}$ denotes only the spin-up component of the
wavefunction Eq. (2) in the spinor representation.
Eq. (21) is derived from the following Anomalous Green's function$^{16}$
\begin{equation}
F({\br}, {\br'}, \omega) = \int \Delta({\bl})G^{\upa}_{\omega}({\br}, {\bl})
G^{\dna}_{-\omega}({\br'}, {\bl})d{\bl}.
\end{equation}

However, Eq. (24) does not satisfy the homogeneity condition after
averaging out the impurity positions, that is,
\begin{equation}
\overline{F({\br}, {\br'}, \omega)}^{imp} \not= 
\overline{F({\br}-{\br'}, \omega)}^{imp}. 
\end{equation}
Substituting Eqs. (22) and (23) into Eq. (24) we find extra pairing terms
such as
\begin{eqnarray}
\overline{\psi_{\vk\upa}({\br})\psi_{\vk'\dna}({\br'})}^{imp} 
 &=& e^{i(\vk\cdot{\br}+\vk'\cdot{\br '})} [1 + O(J^{2}) + \cdots ]\nonumber\\
 &\not=& f({\br}-{\br'}).
\end{eqnarray}
Even if we assume the (incorrect) constant pair potential, we can not
eliminate  the extra pairing between $\psi_{\vk\upa}$ and $\psi_{\vk'\dna}$
because up spin and down spin electrons feel different potentials.
Notice that
\begin{equation}
\int \psi_{\vk\upa}^{*}({\bl})\psi_{\vk'\dna}({\bl})d{\bl}\not= \delta_{\vk\vk'}.
\end{equation}
In fact, the inclusion of the extra pairing has been claimed the origin of the 
so-called pair-breaking of the magnetic impurities.$^{15, 17}$
However the extra pairing terms violate the physical constraint of the
Anomalous Green's function.

The remedy is to incorporate the pairing constraint derived from the
Anomalous Green's function into the self-consistency equation.
The revised self-consistency equation is
\begin{equation}
\Delta({\br}) = VT\sum_{\omega}\int \Delta({\bl})\{G^{\upa}_{\omega}({\br,\bl})
G^{\dna}_{-\omega}({\br,\bl})\}^{\rm P}d\bl,
\end{equation}
where superscript P denotes the pairing constraint which dictates
pairing between $\psi_{\vk\upa}$ and $\psi_{-\vk\dna}$.
Notice that Eq. (28) is nothing but another form of the BCS gap equation,
\begin{equation}
\Delta_{\vk}=\sum_{\vk'}V_{\vk\vk'}{\Delta_{\vk'}\over 2\esn_{\vk'}}tanh
({\esn_{\vk'}\over 2T}),
\end{equation}
where
\begin{equation}
\Delta_{\vk}=\int\psi_{\vk\upa}^{*}({\br})\psi_{-\vk\dna}^{*}({\br})
\Delta({\br}) d{\br},
\end{equation}
and
\begin{equation}
V_{\vk\vk'}=V\int \psi_{\vk'\upa}^{*}({\br})\psi_{-\vk'\dna}^{*}({\br})
\psi_{-\vk\dna}({\br})\psi_{\vk\upa}({\br})d{\br}.
\end{equation}

\section*{5. Conclusion}
\vspace{1pc}
Using the theory by Kim and Overhauser, we have studied 
the compensation of magnetic impurity effect in superconductors as a 
consequence of radiation damage.
Good agreement with the experimental data was found.
We also showed that Gor'kov's formalism with pairing constraint
derived from the Anomalous Green's function gives rise to the KO
theory.

\vspace{3pc}
\centerline{\bf Acknowledgments}
\vspace{1pc}
This work was supported by the Korea Research Foundation through
Domestic Postdoctoral program. 
YJK is grateful to A. W. Overhauser for discussions.
YJK also acknowledges the supports by the Brainpool project
of KOSEF and the MOST.

\vskip 2pc

\vfill\eject

\centerline      {\bf References} 
\vskip 2pc\hb
1. Y.-J. Kim and A. W. Overhauser, Phys. Rev. B{\bf 49}, 15779 (1994).\hb
2. A. A. Abrikosov and L. P. Gor'kov, Sov. Phys. JETP {\bf 12}, 1243 (1961).\hb
3. M. F. Merriam, S. H. Liu, and D. P. Seraphim, Phys. Rev. {\bf 136}, A17 (1964).\hb 
4. G. Boato, M. Bugo, and C. Rizzuto, Phys. Rev. {\bf 148}, 353 (1966).\hb
5. G. Boato and C. Rizzuto, (to be published), (referenced in A. J. Heeger, (1969), in

Solid State Physics {\bf 23}, eds. F. Seitz, D. Turnbull and H. Ehrenreich 
(Academic

 Press, New York), p. 409)  \hb
6. W. Bauriedl and G. Heim, Z. Phys. B{\bf 26}, 29 (1977)\hb
7. M. Hitzfeld and G. Heim, Sol. Stat. Comm. {\bf 29}, 93 (1979).\hb
8. A. Hofmann, W. Bauriedl, and P. Ziemann, Z. Phys. B{\bf 46}, 117 (1982).\hb
9. Y.-J. Kim, Mod. Phys. Lett. B {\bf 10}, 555 (1996).\hb
10. Y.-J. Kim, Int. J. Mod. Phys. B {\bf 11}, 1751 (1997).\hb
11. P. W. Anderson, J. Phys. Chem. Solids {\bf 11}, 26 (1959).\hb
12. J. Bardeen, L. N. Cooper, and J. R. Schrieffer, Phys. Rev. {\bf 108}, 1175 (1957).\hb
13. P. W. Andrson and P. Morel, Phys. Rev. {\bf 123}, 1911 (1961). \hb
14. A. Hofmann, P. Ziemann, and W. Buckel, Nucl. Instrum. Methods {\bf 182}/{\bf 183}, 

943 (1981).\hb
15. P. G. de Gennes and G. Sarma, J. Appl. Phys. {\bf 34}, 1380 (1963)\hb
16. A. A. Abrikosov, L. P. Gor'kov, and I. E. Dzyaloshinski, {\sl Methods of
Quantum Field 

Theory in Statistical Physics} (Prentice-Hall, Englewood, NJ, 1963), 
Eq. (38.3).\hb
17. P. G. de Gennes, {\sl Superconductivity of Metals and Alloys} (Benjamin, New York, 1966) 

P. 266.\hb

\vfill\eject

\centerline{\bf Figure Caption}
\vskip 1pc

{\bf Fig. 1.} Superconducting transition temperature $T_{c}$ of In (open 
symbols) and In-Mn (closed symbols) vs Ar fluence. Data are due to
Hofmann, Bauriedl, and Ziemann, Ref. 8. $1/ \tau_{s}$ was adjusted
in the theoretical curve (lower curve) so that $T_{co}=1.15K$ without 
irradiation. 

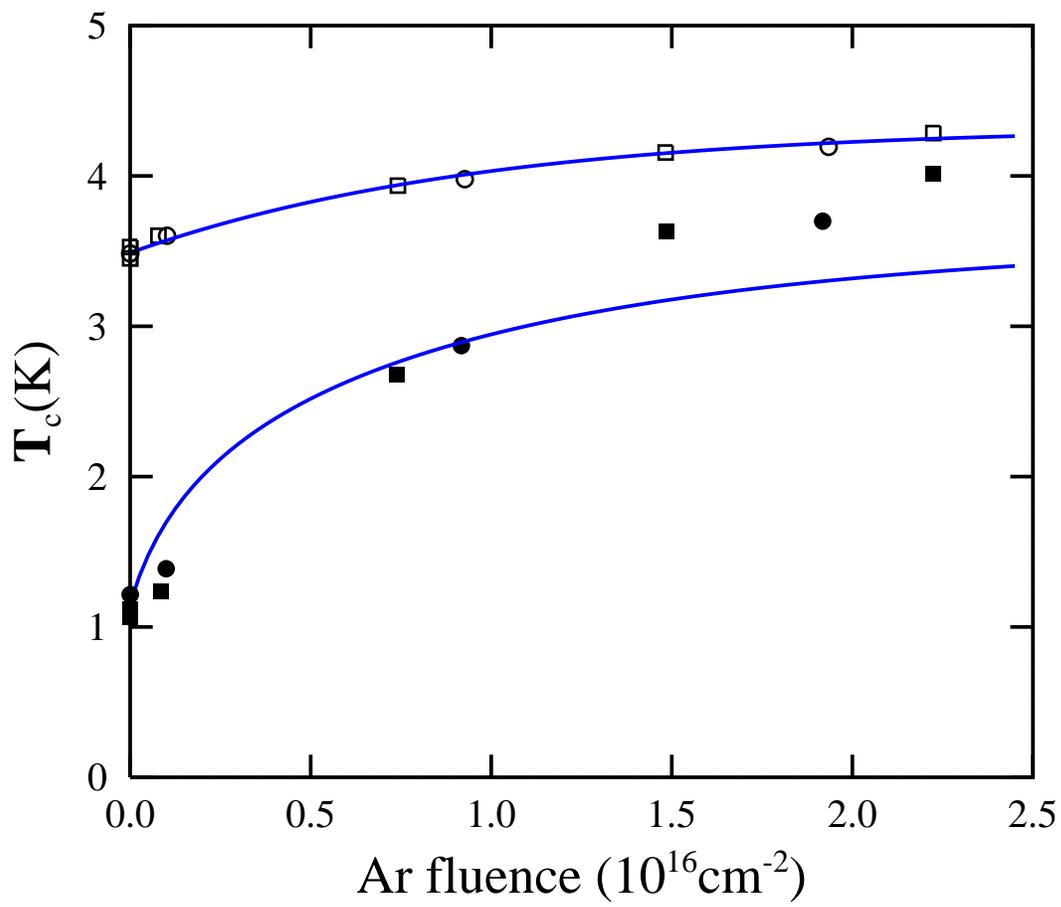
\end{document}